\documentclass[prl,showpacs,superscriptaddress,twocolumn]{revtex4-1}

\usepackage{graphicx}
\usepackage{amsmath}
\usepackage{amssymb}
\usepackage{hyperref}\hypersetup{colorlinks=true,linkcolor=blue,citecolor=blue,pdfstartview=FitH}
\usepackage{bm}\renewcommand{\mathbf}{\bm}

\newcommand{\figwidth}{0.95\columnwidth}

\DeclareMathOperator{\Tr}{Tr}

\begin{document}
\title{Quantum Quench and Prethermalization Dynamics in \\
A Two-Dimensional Fermi Gas
with Long-range Interactions}
\author{N.~Nessi}
\author{A.~Iucci}
\affiliation{Instituto de F\'{\i}sica La Plata (IFLP) - CONICET and Departamento de F\'{\i}sica,\\
Universidad Nacional de La Plata, CC 67, 1900 La Plata, Argentina}
\author{M.~A.~Cazalilla}
\affiliation{Deparment of Physics, National Tsing Hua University, and National Center for Theoretical Sciences (NCTS), Hsinchu City, Taiwan.}

\begin{abstract}
We study the effect of suddenly turning on a long-range interaction in a spinless Fermi gas  in two dimensions. The short to intermediate time dynamics is obtained using the method of bosonization of the Fermi surface. This allow to calculate the full space-time dependence of the non-equilibrium fermion density matrix as well as the evolution of the quasiparticle residue after the quench. It is thus found that the asymptotic state predicted by bosonization is consistent with the prethermalized state. From the bosonized representation, we explicitly construct the Generalized Gibbs Ensamble that describes the prethermalized state. A protocol to perform an interaction quantum quench in a dipolar gas of Erbium atoms is also described.
\end{abstract}

\maketitle

Prethermalization~\cite{berges04_prethermalization_idea} is a
quasi-stationary state that precedes the true thermalization
of a system that has been driven out of equilibrium.
It is often characterized by a fast loss of memory of the initial conditions   due to \emph{dephasing}. The understanding of how such a state emerges
is relevant to  fields  as diverse as heavy ion-collisions~\cite{berges04_prethermalization_idea,berges04_nonequilibrium_field_theory} and non-equilibrium phenomena  in condensed matter and atomic
systems~\cite{polkovnikov11_nonequilibrium_dynamics,cazalilla11_1D_bosons}.  Currently, the subject is attracting much attention in
connection with the thermalization time scales
of ultracold atomic systems
undergoing a quantum quench~\cite{kinoshita06_non_thermalization}.
These systems are ideal for the study of thermalization
of many-body systems
 as they can remain  quantum coherent and isolated from their  environment
for rather long times~\cite{cazalilla11_1D_bosons}.

 Indeed, prethermalization has recently been experimentally observed
in a two-stage equilibration process of two coupled one-dimensional Bose gases~\cite{gring12_pre-thermalization_isolated_bose_gas, kitagawa11_prethermalization_distribution_noise}.  Theoretically, it  has been also reported in studies of   interaction quantum quenches in the Fermi Hubbard model~\cite{moeckel08_quench_hubbard_high_d,eckstein09_thermalization_quench_fermi_hubbard}.  Several groups  have pointed out~\cite{gring12_pre-thermalization_isolated_bose_gas,kollar11_gge_pretherm}  that the prethermalized state  may be  describable  in terms of a generalized Gibbs ensemble (GGE)~\cite{rigol07_generalized_gibbs_hcbosons,cazalilla06_quench_LL,cazalilla12_thermalization_correlations}. The latter
applies to the description of the the long-time behavior of integrable models. The claim is that, even if the model  is non-integrable and ultimately
thermalizes~\cite{rigol08_mechanism_thermalization},  during the first stage
of a  two-stage equilibration process, the system essentially behaves
as integrable. However, the connection between the GGE and prethermalization
 is still poorly understood. For instance, in the studies of
 Refs.~\onlinecite{moeckel08_quench_hubbard_high_d, eckstein09_thermalization_quench_fermi_hubbard},
 the set of integrals of motion required for the
 construction of the GGE has not been  identified.

\begin{figure}[b]
  \includegraphics[width=\figwidth]{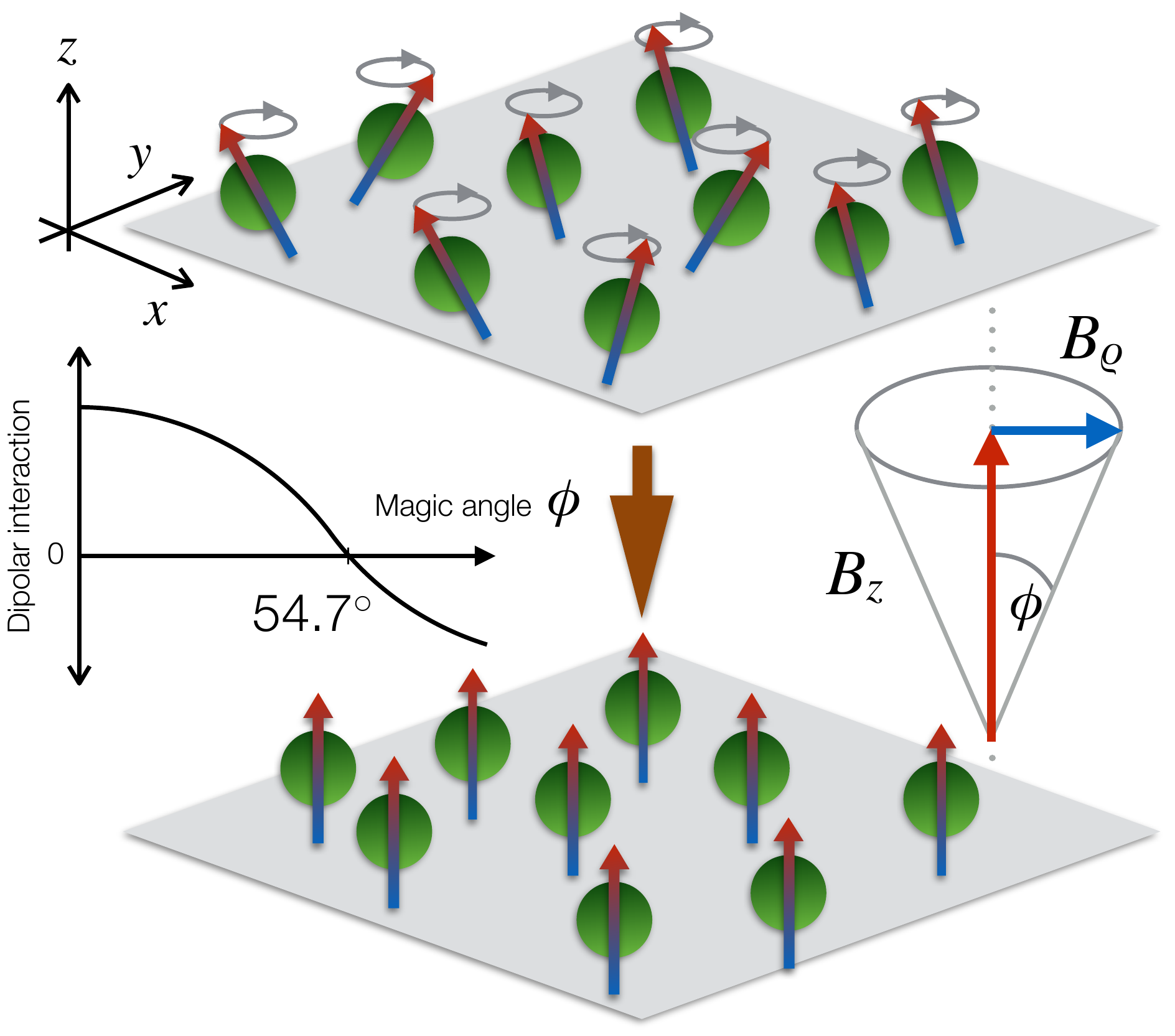}\\
  \caption{(Color Online)
Interaction Quench Protocol: Initially a two dimensional dipolar Fermi gas of, e.g., $^{167}$Er atoms is subject to a DC magnetic field, perpendicular to the plane of atomic motion ($B_z$) and an in plane AC field ($B_{\rho}$). The relative magnitude of the two is fixed so that the atomic dipole moments precess around the $z$ axis with a precession angle equal to the magic angle for which the long range dipolar interaction between the atoms averages to zero and the non-interacting initial state is realized (upper panel). At $t=0$ the DC field is turned off (lower panel).
  }\label{fig:protocol}
\end{figure}

  The two-stage thermalization scenario of a non-integrable system
assumes that there is clear separation of time scales. The first stage is dominated by
the inertial response of the system followed by a rapid dephasing,
whereas the second stage is dominated by
inelastic particle collisions.
The latter  drive  the system to its ultimate
stationary state, which often correspond to a thermal state. A simple model to
understand this is the response of a conductor
to a quench in the electric field, $E(t) = E \theta(t)$, where $\theta(x)$ is the Heaviside
function. The time dependence of the
electric current can easily  obtained within the Drude model
$j(t) = \frac{e^2 \rho_0 E\tau}{m}(1 - e^{-t/\tau})$, where $\rho_0, m$ are the electron density and mass, respectively, and $\tau$ is the mean collision time.
For $t\ll \tau$, the response of the system is inertial $j(t) \simeq \frac{e^2 \rho_0 E}{m} t$.
However, for $t \sim \tau$, it crosses over to the stationary state where
$j(t \gg \tau)\approx  \frac{e^2 \rho_0 E\tau}{m} =$ const. This analogy is actually quite relevant to the study reported below, as the inertial regime
is controlled by the so-called \emph{anomaly}, which determines the quantum
commutator of the current and density operators. Indeed,
such ``anomalous" commutators also determine  the inertial response of a Fermi gas
following an interaction quench.

For an interaction quench, as pointed out in Ref.~\cite{moeckel08_quench_hubbard_high_d},
the typical inter-particle  collision time~\footnote{For weak interactions and on dimensional grounds, we expect the collision time
to be $\sim f_0^{-4} N(0)^{-3}$, where $f_0$ is the interaction strength and $N(0)$ is the density of states at the Fermi surface, while the prethermalization regime arises for times of the order $f_0^{-2} N(0)^{-1}$.} is the characteristic
time scale at which  the system crosses over from stage one to stage two.
In this work, we take this separation for granted
and construct a non-perturbative description of the
first stage of the thermalization of a Fermi gas after a quench
in which the interaction is
suddenly switched on. Unlike earlier
studies~\cite{moeckel08_quench_hubbard_high_d,eckstein09_thermalization_quench_fermi_hubbard},
the fermions are assumed to be spinless and the interaction
long-ranged. This type of quench can be realized
experimentally  (see Fig.~\ref{fig:protocol})
in  atomic quantum degenerate dipolar Fermi gases,
which have recently become available~\cite{lu12_erbium_degenerate}.

On the theory side, our approach also allows us
to identify the integrals of motion  that constrain
the dynamics leading to the prethermal state.
This is achieved by first truncating the interaction term
of the Hamiltonian and retaining only
forward-scattering  interactions between the fermions.
Interactions leading to inelastic inter-particle collisions, which, according to what
has been described above, set in at longer times are therefore neglected.
This approximation is akin to Fermi-liquid theory (FLT) in the equilibrium case.
The major difference is that our truncated Hamiltonian is written in terms of the
\emph{bare} interaction, instead of a renormalized interaction between Landau
quasi-particles. Nevertheless, as in Landau FLT~\cite{houghton93_bosonization_high_d}, the truncated Hamiltonian exhibits a Gauge $\mathrm{U}(1)_k^{\infty}$ symmetry in momentum space, which renders the dynamics exactly soluble.

 Next, we solve for the dynamics of the truncated Hamiltonian
  by performing a  change to a basis where the particle-hole
excitations of the Fermi gas are described by a set of a harmonic oscillator modes
(Tomonaga bosons) at each point of the Fermi surface (FS)~\cite{houghton93_bosonization_high_d,castroneto94_bosonization_low_energy_fermi_liquids}.
This is possible because the density operator, when projected to the
neighborhood of the FS, obeys a set of anomalous commutation relations.
The approach, known as FS bosonization~\cite{houghton93_bosonization_high_d,castroneto94_bosonization_low_energy_fermi_liquids},
allows to threat the truncated Hamiltonian non-perturbatively.
The technique  focuses on the low-energy FS
degrees of freedom, which are most affected at  short times after the quench.  This is because, for a long-ranged interaction,
small-momentum interactions are dominant,
and thus particle-holes in the neighborhood of the FS are created
at short times.

 Furthermore, the conserved quantities of the truncated
Hamiltonian  can be identified as the occupation operators of the Tomonaga bosons.
Our approach  also allows  us  to
unveil the close relationship linking prethermalization, dephasing dynamics, and GGE for a generic Fermi liquid  of \emph{finite} dimensionality, which is very similar to the phenomena observed in exactly solvable models in one space dimension (1D)~\cite{cazalilla12_thermalization_correlations}.

Indeed, for 1D systems recent numerical studies~\cite{karrasch12_ll_universality_quench} have shown that the description  provided  in Refs.~\onlinecite{cazalilla06_quench_LL} is fairly
accurate and universal and that even the dynamics at rather long times can
be described by the GGE although mode coupling has been shown to lead to thermalization~\cite{mitra11_quench_mode_coupling}. In the two dimensional case, we find similarities and also important differences with 1D. In fact, the picture that emerges from the bosonization treatment of the two-dimensional Fermi liquid suggest that the first stage following the quench can be regarded as a collective  \emph{non-equilibrium} quasi-particle dressing by the interactions of the non-interacting fermions (or, for that matter, the bare quasi-particles) that are the fermionic excitations of the initial Hamiltonian.

\begin{figure}[t]
  \includegraphics[width=\figwidth]{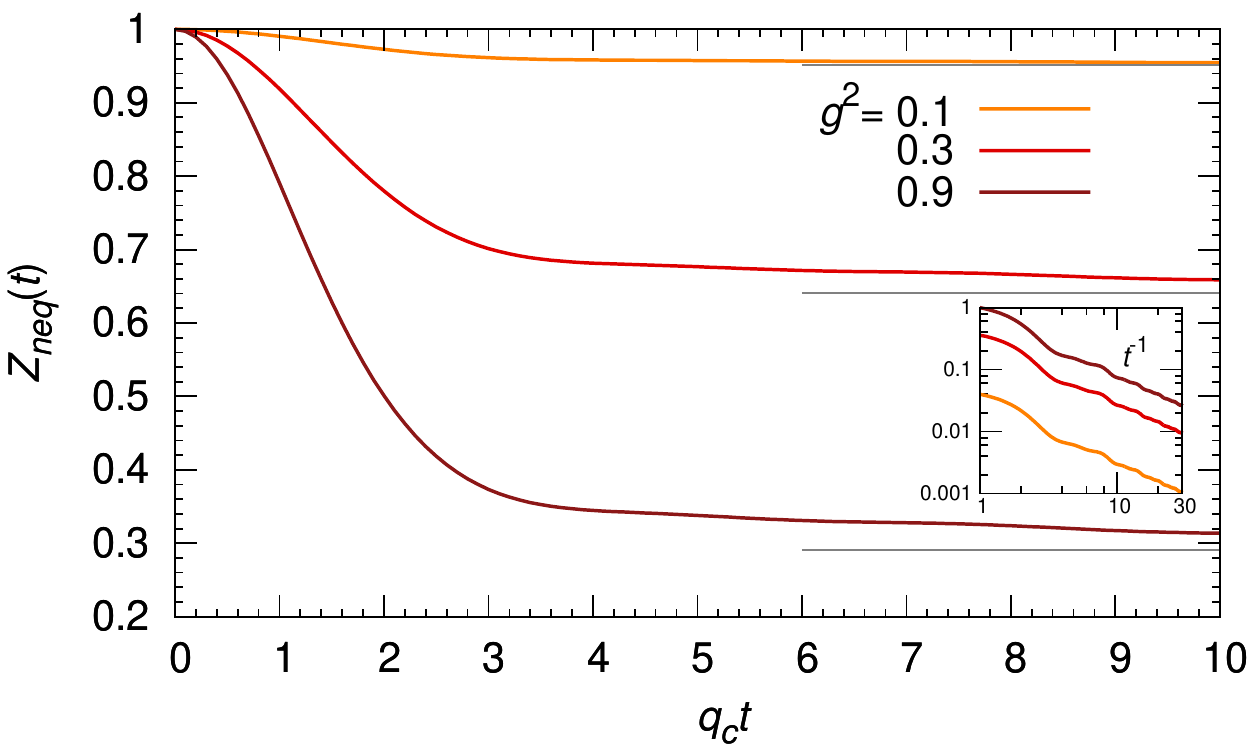}\\
  \caption{(Color Online) The discontinuity in the zero-temperature
  momentum distribution at the Fermi momentum,
   $Z^{\mathrm{neq}}(t)$, exhibits a Gaussian decay at short times. Asymptotically, it saturates to a finite value $Z^{\mathrm{st}}\simeq (Z^{\mathrm{eq}})^{2}$ (horizontal lines) in the stationary (prethermalized) state. The effective interaction strength is $g^{2}=\frac{4f_0^{2}q_{c}N(0)}{(2\pi)^3}$.   Inset: $\ln(Z^{\mathrm{neq}}(t)/Z^{\mathrm{st}})$ showing the asymptotic behavior $\sim \exp[{-t^{-1}}]$.}\label{fig:z}
\end{figure}

The  \emph{non-equilibrium} dressing is best displayed by
the time-evolution of the discontinuity at the Fermi momentum of the (zero-temperature) momentum distribution, $Z^{\mathrm{neq}}(t)$ (Fig.~\ref{fig:z}). The latter decreases from its initial  unity value and asymptotically (i.e. for $t \to +\infty$) reaches a value
smaller  than the equilibrium value (cf. Fig.~\ref{fig:z}), which, interestingly, can be obtained using the GGE (see below for details). A similar behavior of $Z^{\mathrm{neq}}(t)$ has been observed in interaction quenches in 1D~\cite{cazalilla06_quench_LL,karrasch12_ll_universality_quench}, but a major difference with the present case is that $Z_{1D}^{\mathrm{neq}}(t\to +\infty) \to 0$, indicating the non-equilibrium dressing leads to a complete destruction of the fermionic quasi-particles in the system. The plateau exhibited by $Z^{\mathrm{neq}}(t)$ after the initial rapid (Gaussian-like) decay can be regarded as a consequence of the system reaching a state that relaxes very slowly. We argue that this  plateau is a signature of a prethermalized state. Interestingly, a similar behavior has been observed previously in studies of interaction quenches in the Hubbard model~\cite{moeckel08_quench_hubbard_high_d,eckstein09_thermalization_quench_fermi_hubbard}.

The inclusion of inelastic collisions breaks integrability and violates the Gauge U$(1)^{\infty}_{k}$ symmetry. A complete relaxation to thermal equilibrium is thus expected to occur at long times due to the decay of quasiparticles (and quasi-holes) excited away from the FS.
The latter phenomenon is controlled only by the conservation laws that are respected by the
true fermion Hamiltonian, which differs from the truncated FLT Hamiltonian because it also describes inelastic collisions.

In what follows, we outline the most important details of the derivation of the results discussed above and displayed in Fig.~\ref{fig:z}. We refer the reader to the supplementary material for further details. Let us consider a system of  spinless fermions in two dimensions. We study a quantum quench in which an interaction of (spatial) range
$\sim q^{-1}_c \gg k^{-1}_F$ is \emph{suddenly} switched on  at time $t = 0$,   ($k_F = \sqrt{4 \pi \rho_0}$ is the Fermi momentum and $\rho_0 = N/V$ the mean fermion density). For simplicity, we assume a circular FS of radius $k_{F}$. In order to describe the excitations that are produced shortly after the quantum quench, we restrict our attention  to a description  of the processes inside a narrow shell of thickness $\lambda > q_c$ around the FS ($\lambda\ll k_{F}$). In addition, we divide the FS into $N$ patches of dimensions $\Lambda$ along the FS that are labeled by the vector $\mathbf{S}$. We start from a generic Hamiltonian for spinless fermions with two-body interactions $H=H_0+H_{\mathrm{int}}$, with $H_0=\sum_{\mathbf{k}}\epsilon(\mathbf{k})c^{\dagger}_{\mathbf{k}}c_{\mathbf{k}}$ and $H_{\mathrm{int}}=\frac{1}{2V}\sum_{\mathbf{k}_1,\mathbf{k}_2,\mathbf{q}}f(q)J_{\mathbf{k}_1}(\mathbf{q})J_{\mathbf{k}_2}(-\mathbf{q})$, where $\{c_{\mathbf{k}},c^{\dagger}_{\mathbf{k}}\}$ are the fermionic field operators,  $\epsilon(\mathbf{k})$ is the dispersion relation,  $f(q)$  is the \emph{bare} (isotropic) two-body interaction that is switched on at $t = 0$ and $J_{\mathbf{k}}(\mathbf{q})=c^{\dagger}_{\mathbf{k}+\mathbf{q}}c_{\mathbf{k}}$ create particle-hole excitations. Using the patch construction we truncate the Hamiltonian by retaining those terms that describe
forward and exchange scattering processes between patches. The latter conserve the
number of fermions within each patch, which in the limit of a large number of patches becomes a $U(1)^{\infty}_k$ Gauge symmetry~\cite{houghton93_bosonization_high_d}. Additionally, this sectorization allows to linearize the dispersion relation within each patch, provided that the patch size $\Lambda$ is small enough compared to the scale in which the FS changes its shape (in our case it is enough to set $\Lambda\ll k_{F}$). The truncated Hamiltonian can be written in the form~\cite{houghton93_bosonization_high_d}:
\begin{equation}\label{eq:bos_ham_main}
H=\frac{1}{2}\sum_{\mathbf{S},\mathbf{T},\mathbf{q}}J_{\mathbf{S}}(\mathbf{q})\left(\frac{v_F}{\Omega} \delta_{\mathbf{S},\mathbf{T}} +\frac{f(q)}{V}\right)J_{\mathbf{T}}(-\mathbf{q}),
\end{equation}
where $v_F=\vert\nabla_{\mathbf{k}}\epsilon(\mathbf{k})\vert_{\vert\mathbf{k}\vert=k_F}$ is the Fermi velocity and $J_{\mathbf{S}}(\mathbf{q})$ are the Fourier components of the density
operator corresponding to momenta within the patch $\mathbf{S}$~\cite{supplementary}. Thus, these
operators create  particle-hole pairs near the FS with momentum $\mathbf{q}$ within a distance $\Lambda$ of the direction $\mathbf{S}$, $\Omega=\Lambda\frac{V}{(2\pi)^2}$ ($V$ is the area of the system). Key to  bosonization is the observation that the fermonic currents $J_{\mathbf{S}}(\mathbf{q})$ obey a set of anomalous commutation relations, which becomes an approximate bosonic algebra provided that $\Lambda\gg\lambda$~\cite{houghton93_bosonization_high_d,castroneto94_bosonization_low_energy_fermi_liquids}. This allows for a representation of the currents in terms of Tomonaga bosons which  renders the Hamiltonian quadratic.
Higher order (i.e. anaharmonic) corrections to~\eqref{eq:bos_ham_main} such as those arising from  the curvature of the FS and
the  fermion dispersion, introduce interactions between the Tomonaga bosons~\cite{kopietz99_book_bosonization_high_D}.
These terms break the infinite set of conservation laws that make the
above Hamiltonian exactly solvable~\cite{houghton93_bosonization_high_d}.
As discussed above, we neglect those terms as we are interested in the short-time prethermalization dynamics.

The system is driven out of equilibrium by suddenly turning the interactions, $H_{\mathrm{int}}$, at $t =0$. This amounts to initially prepare the system in the ground state $\vert\Psi_0\rangle$ of $H_0$ and letting it subsequently evolve under $H = H_0 + H_{\mathrm{int}}$.  In order to calculate correlation functions, we must solve the Heisenberg equations of motion for the bosonic operators. This is equivalent to diagonalizing the bilinear Hamiltonian, which is a non-trivial task because all patches are coupled  by the interaction. The explicit diagonalization can be bypassed by noticing that the solution to the equations of motion can be written exclusively in terms of the \emph{equilibrium} retarded correlation functions between the bosonic fields,  which follows from the bilinear form of the Hamiltonian. Those retarded correlations can be calculated exactly~\cite{houghton93_bosonization_high_d}.

 We first discuss the asymptotic state of the system at long times (formally, for $t \to +\infty$), and argue that its  properties are consistent with those of a prethermalized state. To this end, we consider the non-equilibrium density matrix $\mathcal{G}^{\mathrm{neq}}_{\mathbf{S}}(\mathbf{x},t)
=\langle\Psi_0\vert e^{iHt}\psi^{\dagger}_{\mathbf{S}}(\mathbf{x})\psi_{\mathbf{S}}(\mathbf{0})e^{iHt}\vert\Psi_0\rangle$.
The calculation can be carried out by means of the bosonic representation of the fermionic field~\cite{houghton93_bosonization_high_d,castroneto94_bosonization_low_energy_fermi_liquids}. We find $\mathcal{G}^{\mathrm{neq}}_{\mathbf{S}}(\mathbf{x},t)=\mathcal{G}^{0}_{\mathbf{S}}(\mathbf{x})Z^{\mathrm{neq}}_{\mathbf{S}}(\mathbf{x},t)$, where $\mathcal{G}^{0}_{\mathbf{S}}(\mathbf{x})$ is the in-patch free correlator (the initial condition) and
\begin{multline}\label{eq:z_ret}
Z^{\mathrm{neq}}_{\mathbf{S}}(\mathbf{x},t)=\exp\Bigg[\frac{1}{\Omega}\sum_{\mathbf{q},\mathbf{\hat{n}_S}\cdot\mathbf{q}>0,\mathbf{T}}\frac{\vert\tilde{G}^{\mathrm{ret,eq}}_{\mathbf{S}\mathbf{T}}(\mathbf{q},t)\vert^2}{\mathbf{\hat{n}_S}\cdot\mathbf{q}}\\
\times2(\cos(\mathbf{q}\cdot\mathbf{x})-1)\Bigg],
\end{multline}
where $\tilde{G}^{\mathrm{ret,eq}}_{\mathbf{S}\mathbf{T}}(\mathbf{q},t)$ is the anomalous retarded correlation function of the Tomonaga bosons and $\mathbf{\hat{n}_S}$ is the unit normal to the FS at patch $\mathbf{S}$. In what follows we shall assume that $f(q)=f_0\,F(q)$, where $F(q)$ decays quickly for $q>q_c$ and $f_0$ is a small parameter. Although it is possible to obtain an expression for $\tilde{G}^{\mathrm{ret,eq}}$ valid at all orders of the interaction strength, to produce analytically tractable expressions, we shall keep only the leading order.

Next, we relate the stationary state and the ground state correlations. Thus,
considering the correlation function $\mathcal{G}^{\mathrm{eq}}(\mathbf{x})=\langle\Psi\vert\psi^{\dagger}(\mathbf{x})\psi(\mathbf{0})\vert\Psi\rangle$ ($\vert\Psi\rangle$ is the ground state of $H$), we define $Z^{\mathrm{eq}}(x) = \mathcal{G}^{\mathrm{eq}}(x)/\mathcal{G}^{0}(x)$. Furthermore, due to dephasing, the oscillatory part of $\tilde{G}^{\mathrm{ret,eq}}_{\mathbf{S}\mathbf{T}}(\mathbf{q},t)$  drops out for $t\rightarrow\infty$~\cite{supplementary}. We then find that
\begin{equation}\label{eq:z_neq}
\lim_{t\rightarrow\infty} \ln\left[Z^{\mathrm{neq}}(x,t)\right]=2\ln\left[Z^{\mathrm{eq}}(x)\right]+\mathcal{O}(f_{0}^{3}),
\end{equation}
where $Z^{\mathrm{neq}}(x,t) = \mathcal{G}^{\mathrm{neq}}(x,t)/\mathcal{G}^0(x)$. Note that, owing to rotation invariance, both  $Z^{\mathrm{eq}}$ and $Z^{\mathrm{neq}}$  depend only on $x=\vert\mathbf{x}\vert$.
Taking  $x\rightarrow\infty$, Eq. (\ref{eq:z_neq})  leads to a relation between the discontinuity at $k_F$ in the momentum distribution of the stationary and ground states:  $Z^{\mathrm{neq}}\simeq(Z^{\mathrm{eq}})^2$.

At this point, we make contact with previous work. Exponentiating Eq. (\ref{eq:z_neq}) and further developing in powers of the interaction strength, it is possible to obtain a simple relation between the stationary state ($n^{\mathrm{st}}(k)$) and ground state ($n^{\mathrm{eq}}(k)$) momentum distributions,
$2\left[n^{\mathrm{eq}}(k)-n^{\mathrm{eq}}_{0}(k)\right]=n^{\mathrm{st}}(k)-n^{\mathrm{eq}}_{0}(k)+\mathcal{O}(f_{0}^{3})$,
where $n^{\mathrm{eq}}_{0}(k)  =\theta(k_F-k)$. A similar result was obtained  in Ref.~\onlinecite{moeckel08_quench_hubbard_high_d} for the Hubbard model under
very different assumptions (a short range interaction between spinful fermions). In particular,
this relation  implies that, to the lowest order in the interaction strength, all the energy injected into the system by the quench, $E_\mathrm{ex} = E_{\mathrm{neq}} - E_{\mathrm{eq}}$, where $E_{\mathrm{neq}} =  \langle \Psi_0| H | \Psi_0 \rangle =0$ and $E_{\mathrm{eq}} = \langle \Psi| H | \Psi \rangle $, is transformed into kinetic
energy in the stationary
(prethermalized) state~\cite{moeckel08_quench_hubbard_high_d}. As to the role of higher order corrections to this picture, we find by direct diagonalization  of the bosonic Hamiltonian that the quench also excites the collective mode of the fermionic system and that the excitation energy is \emph{almost} completely transferred for long times after the quench into kinetic energy, i.e., $K_{\infty}=K_\mathrm{GS}+E_\mathrm{ex}-\delta U$, where $K_{\infty}$ ($K_\mathrm{GS}=\langle \Psi| H_0 | \Psi \rangle$) is the kinetic energy in the steady (ground) state and $\delta U$ is a correction arising from the collective mode that is of order $f_0^4$ for weak interactions, and therefore not seen in perturbative calculations at the lowest orders.

Next, we discuss the statistical description of the stationary prethermalized state. To this end,
we recall that the truncated Hamiltonian is a (bosonic) bilinear  and, consequently, that
dephasing implies  that \emph{all} correlations in the steady state are described by a
GGE~\cite{cazalilla12_thermalization_correlations}. If we denote with $\{\alpha_{l}(\mathbf{q}),\alpha^{\dagger}_{l}(\mathbf{q})\}$ the bosonic basis that diagonalizes the Hamiltonian~\eqref{eq:bos_ham_main}, the GGE density matrix can be written as
\begin{equation}\label{eq:gge}
\rho_{\mathrm{GGE}}=\frac{1}{Z_{\mathrm{GGE}}}\exp\left[\sum_{l,\mathbf{q}}\lambda_{l}(\mathbf{q})I_{l}(\mathbf{q})\right],
\end{equation}
where  $I_{l}(\mathbf{q})=\alpha^{\dagger}_{l}(\mathbf{q})\alpha_{l}(\mathbf{q})$ are the conserved quantities, $Z_{\mathrm{GGE}}=\Tr[\rho_{\mathrm{GGE}}]$ and the Lagrange multipliers $\lambda_{l}(\mathbf{q})$ are obtained from the initial conditions, $\langle I_{l}(\mathbf{q})\rangle_{t=0}=\langle\Psi_0\vert I_{l}(\mathbf{q})\vert\Psi_0\rangle=\Tr  \left[\rho_{\mathrm{gG}} I_{l}(\mathbf{q})\right]$. We also have explicitly checked that the density matrix~\eqref{eq:gge} reproduces all the studied quantities in the prethermalized state. It is also worth noting that the conserved quantities can be easily refermionized, at least formally. Using the matrix transformation that diagonalizes the Hamiltonian they can be expressed as linear combination of products of two patch densities.

We finally take up the dynamics at finite $t$. An isotropic horizon effect in the correlations~\cite{calabrese06_quench_CFT} arises at the lowest order in the expansion of $\ln[Z^{\mathrm{neq}}_{\mathbf{S}}(\mathbf{x},t)]$ in powers of the interaction strength. In fact, it can be shown~\cite{supplementary} that in the spatio-temporal region defined by $\vert\mathbf{x}\vert\gg 2v_Ft$ we can approximately neglect the spatial dependence of $Z_{\mathbf{S}}^{\mathrm{neq}}(\mathbf{x},t)$ and, with it, the patch index. Outside the light cone, $\vert\mathbf{x}\vert\gg 2v_Ft$, the interaction correction is therefore approximately the same for all patches: $Z^{\mathrm{neq}}_{\mathbf{S}}(\mathbf{x},t)\approx Z^{\mathrm{neq}}(t)$ and the full correlation function thus reads $\mathcal{G}^{\mathrm{neq}}(x,t)\approx\mathcal{G}^{\mathrm{0}}(x)Z^{\mathrm{neq}}(t)$, i.e., the correlations retain the same spatial dependence as in the initial state up to a time-dependent prefactor. This factor defines the time-dependent quasiparticle residue that is analyzed below. In the opposite limit, $\vert\mathbf{x}\vert \ll 2v_Ft$, we can neglect the temporal dependence and the steady state correlations dominate: $Z^{\mathrm{neq}}(x,t)\approx \lim_{t\rightarrow\infty}Z^{\mathrm{neq}}(x,t)$.

We next turn to the dynamics of the discuntinuity of the momentum distribution at $k = k_F$.
For short times $t\ll q^{-1}_c$ we find a Gaussian decay of $Z^{\mathrm{neq}}(t)$ from its initial value of one:
\begin{equation}\label{eq:short_t}
Z^{\mathrm{neq}}(t)=\exp\left[-t^{2}\frac{4N(0)}{(2\pi)^{3}}C\int^{\infty}_{0}dq\,(f(q)q)^2+\mathcal{O}(f_{0}^{3})\right],
\end{equation}
where $N(0)=\frac{k_F}{2\pi}$ is the density of states at the FS and $C$ is an $\mathcal{O}(1)$ constant that stems from the angular integration over the FS. The Gaussian decay at short times is independent of the form of the interaction and it also occurs in 1D~\cite{cazalilla06_quench_LL}. For $t\gg q^{-1}_c$, the form of the interaction is required, and upon choosing $f(q)=f_{0}q^{n}e^{-q/q_c}$, we find:
\begin{equation}
Z^{\mathrm{neq}}(t)\approx Z^{\mathrm{st}}\exp[f_{0}^{2}a_{n}(q_c t)^{-(2n+1)}],
\end{equation}
where $Z^{\mathrm{st}}\simeq(Z^{\mathrm{eq}})^2$ is the stationary-state quasiparticle residue and $a_{n}$ a positive constant.
In Figure \ref{fig:z} we illustrate the dynamics of the quasiparticle residue for different interaction strengths and $n=0$.

To conclude, we have studied the quench dynamics of a Fermi gas with long-range interactions using FS bosonization. We were able to obtain the full space-time dependence of the non-equilibrium density-matrix as well as the evolution of (the zero-temperature) discontinuity of the momentum distribution at $k = k_F$ after the quench. We have shown that prethermalization can be understood as the result of dephasing between the bosonic FS excitations. Furthermore,  the statistical description of the prethermalized state in terms of the generalized Gibbs ensemble (GGE)  has been obtained and the integrals of motion needed for its constuction have been identified as the eigenmodes of the bosonic
FS Hamiltonian.

 Finally, we note that the interaction quench described above can be experimentally realized in a quantum degenerate dipolar Fermi gas of, e.g., $^{167}$Er atoms~\cite{lu12_erbium_degenerate} confined to a 2D trap. The details of the protocol are given in Fig.~\ref{fig:protocol}. Based on the results of this work, we predict that, for a weak to moderate interaction quench, the system will exhibit a prethermalized state. The latter will show up as a quasi-stationary state in a two stage thermalization process: After an initial rapid change, the momentum distribution will settle down in a non-thermal behavior different from the final thermal distribution. However, the kinetic energy should rapidly reach its final value.  In future work, grounded on the results reported here, we shall explore different aspects of the quench dynamics, in particular, the final stage of the thermalization dynamics when terms that break integrability are included.

\acknowledgments

This work was partially supported by CONICET (PIP 0662), ANPCyT (PICT 2010-1907) and UNLP (PID X497), Argentina. MAC acknowledges financial support from NSC a start-up fund from NTHU (Taiwan).

%


\begin{thebibliography}{31}%
\makeatletter
\providecommand \@ifxundefined [1]{%
 \@ifx{#1\undefined}
}%
\providecommand \@ifnum [1]{%
 \ifnum #1\expandafter \@firstoftwo
 \else \expandafter \@secondoftwo
 \fi
}%
\providecommand \@ifx [1]{%
 \ifx #1\expandafter \@firstoftwo
 \else \expandafter \@secondoftwo
 \fi
}%
\providecommand \natexlab [1]{#1}%
\providecommand \enquote  [1]{``#1''}%
\providecommand \bibnamefont  [1]{#1}%
\providecommand \bibfnamefont [1]{#1}%
\providecommand \citenamefont [1]{#1}%
\providecommand \href@noop [0]{\@secondoftwo}%
\providecommand \href [0]{\begingroup \@sanitize@url \@href}%
\providecommand \@href[1]{\@@startlink{#1}\@@href}%
\providecommand \@@href[1]{\endgroup#1\@@endlink}%
\providecommand \@sanitize@url [0]{\catcode `\\12\catcode `\$12\catcode
  `\&12\catcode `\#12\catcode `\^12\catcode `\_12\catcode `\%12\relax}%
\providecommand \@@startlink[1]{}%
\providecommand \@@endlink[0]{}%
\providecommand \url  [0]{\begingroup\@sanitize@url \@url }%
\providecommand \@url [1]{\endgroup\@href {#1}{\urlprefix }}%
\providecommand \urlprefix  [0]{URL }%
\providecommand \Eprint [0]{\href }%
\providecommand \doibase [0]{http://dx.doi.org/}%
\providecommand \selectlanguage [0]{\@gobble}%
\providecommand \bibinfo  [0]{\@secondoftwo}%
\providecommand \bibfield  [0]{\@secondoftwo}%
\providecommand \translation [1]{[#1]}%
\providecommand \BibitemOpen [0]{}%
\providecommand \bibitemStop [0]{}%
\providecommand \bibitemNoStop [0]{.\EOS\space}%
\providecommand \EOS [0]{\spacefactor3000\relax}%
\providecommand \BibitemShut  [1]{\csname bibitem#1\endcsname}%
\let\auto@bib@innerbib\@empty
\bibitem [{\citenamefont {Berges}\ \emph {et~al.}(2004)\citenamefont {Berges},
  \citenamefont {Bors{'a}nyi},\ and\ \citenamefont
  {Wetterich}}]{berges04_prethermalization_idea}%
  \BibitemOpen
  \bibfield  {author} {\bibinfo {author} {\bibfnamefont {J.}~\bibnamefont
  {Berges}}, \bibinfo {author} {\bibfnamefont {S.}~\bibnamefont {Bors{'a}nyi}},
  \ and\ \bibinfo {author} {\bibfnamefont {C.}~\bibnamefont {Wetterich}},\
  }\href@noop {} {\bibfield  {journal} {\bibinfo  {journal} {Phys. Rev. Lett.}\
  }\textbf {\bibinfo {volume} {93}},\ \bibinfo {pages} {143002} (\bibinfo
  {year} {2004})}\BibitemShut {NoStop}%
\bibitem [{\citenamefont
  {Berges}(2004)}]{berges04_nonequilibrium_field_theory}%
  \BibitemOpen
  \bibfield  {author} {\bibinfo {author} {\bibfnamefont {J.}~\bibnamefont
  {Berges}},\ }\href@noop {} {\bibfield  {journal} {\bibinfo  {journal} {AIP
  Conf. Proc.}\ }\textbf {\bibinfo {volume} {3}},\ \bibinfo {pages} {739}
  (\bibinfo {year} {2004})}\BibitemShut {NoStop}%
\bibitem [{\citenamefont {Polkovnikov}\ \emph {et~al.}(2011)\citenamefont
  {Polkovnikov}, \citenamefont {Sengupta}, \citenamefont {Silva},\ and\
  \citenamefont {Vengalattore}}]{polkovnikov11_nonequilibrium_dynamics}%
  \BibitemOpen
  \bibfield  {author} {\bibinfo {author} {\bibfnamefont {A.}~\bibnamefont
  {Polkovnikov}}, \bibinfo {author} {\bibfnamefont {K.}~\bibnamefont
  {Sengupta}}, \bibinfo {author} {\bibfnamefont {A.}~\bibnamefont {Silva}}, \
  and\ \bibinfo {author} {\bibfnamefont {M.}~\bibnamefont {Vengalattore}},\
  }\href@noop {} {\bibfield  {journal} {\bibinfo  {journal} {Rev. Mod. Phys.}\
  }\textbf {\bibinfo {volume} {83}},\ \bibinfo {pages} {863} (\bibinfo {year}
  {2011})}\BibitemShut {NoStop}%
\bibitem [{\citenamefont {Cazalilla}\ \emph {et~al.}(2011)\citenamefont
  {Cazalilla}, \citenamefont {Citro}, \citenamefont {Giamarchi}, \citenamefont
  {Orignac},\ and\ \citenamefont {Rigol}}]{cazalilla11_1D_bosons}%
  \BibitemOpen
  \bibfield  {author} {\bibinfo {author} {\bibfnamefont {M.~A.}\ \bibnamefont
  {Cazalilla}}, \bibinfo {author} {\bibfnamefont {R.}~\bibnamefont {Citro}},
  \bibinfo {author} {\bibfnamefont {T.}~\bibnamefont {Giamarchi}}, \bibinfo
  {author} {\bibfnamefont {E.}~\bibnamefont {Orignac}}, \ and\ \bibinfo
  {author} {\bibfnamefont {M.}~\bibnamefont {Rigol}},\ }\href@noop {}
  {\bibfield  {journal} {\bibinfo  {journal} {Rev. Mod. Phys.}\ }\textbf
  {\bibinfo {volume} {83}},\ \bibinfo {pages} {1405} (\bibinfo {year}
  {2011})}\BibitemShut {NoStop}%
\bibitem [{\citenamefont {Kinoshita}\ \emph {et~al.}(2006)\citenamefont
  {Kinoshita}, \citenamefont {Wenger},\ and\ \citenamefont
  {Weiss}}]{kinoshita06_non_thermalization}%
  \BibitemOpen
  \bibfield  {author} {\bibinfo {author} {\bibfnamefont {T.}~\bibnamefont
  {Kinoshita}}, \bibinfo {author} {\bibfnamefont {T.}~\bibnamefont {Wenger}}, \
  and\ \bibinfo {author} {\bibfnamefont {D.~S.}\ \bibnamefont {Weiss}},\
  }\href@noop {} {\bibfield  {journal} {\bibinfo  {journal} {Nature (London)}\
  }\textbf {\bibinfo {volume} {440}},\ \bibinfo {pages} {900} (\bibinfo {year}
  {2006})}\BibitemShut {NoStop}%
\bibitem [{\citenamefont {Gring}\ \emph {et~al.}(2012)\citenamefont {Gring},
  \citenamefont {Kuhnert}, \citenamefont {Langen}, \citenamefont {Kitagawa},
  \citenamefont {Rauer}, \citenamefont {Schreitl}, \citenamefont {Mazets},
  \citenamefont {Smith}, \citenamefont {Demler},\ and\ \citenamefont
  {Schmiedmayer}}]{gring12_pre-thermalization_isolated_bose_gas}%
  \BibitemOpen
  \bibfield  {author} {\bibinfo {author} {\bibfnamefont {M.}~\bibnamefont
  {Gring}}, \bibinfo {author} {\bibfnamefont {M.}~\bibnamefont {Kuhnert}},
  \bibinfo {author} {\bibfnamefont {T.}~\bibnamefont {Langen}}, \bibinfo
  {author} {\bibfnamefont {T.}~\bibnamefont {Kitagawa}}, \bibinfo {author}
  {\bibfnamefont {B.}~\bibnamefont {Rauer}}, \bibinfo {author} {\bibfnamefont
  {M.}~\bibnamefont {Schreitl}}, \bibinfo {author} {\bibfnamefont
  {I.}~\bibnamefont {Mazets}}, \bibinfo {author} {\bibfnamefont {D.~A.}\
  \bibnamefont {Smith}}, \bibinfo {author} {\bibfnamefont {E.}~\bibnamefont
  {Demler}}, \ and\ \bibinfo {author} {\bibfnamefont {J.}~\bibnamefont
  {Schmiedmayer}},\ }\href@noop {} {\bibfield  {journal} {\bibinfo  {journal}
  {Science}\ }\textbf {\bibinfo {volume} {337}},\ \bibinfo {pages} {1318}
  (\bibinfo {year} {2012})}\BibitemShut {NoStop}%
\bibitem [{\citenamefont {Kitagawa}\ \emph {et~al.}(2011)\citenamefont
  {Kitagawa}, \citenamefont {Imambekov}, \citenamefont {Schmiedmayer},\ and\
  \citenamefont {Demler}}]{kitagawa11_prethermalization_distribution_noise}%
  \BibitemOpen
  \bibfield  {author} {\bibinfo {author} {\bibfnamefont {T.}~\bibnamefont
  {Kitagawa}}, \bibinfo {author} {\bibfnamefont {A.}~\bibnamefont {Imambekov}},
  \bibinfo {author} {\bibfnamefont {J.}~\bibnamefont {Schmiedmayer}}, \ and\
  \bibinfo {author} {\bibfnamefont {E.}~\bibnamefont {Demler}},\ }\href@noop {}
  {\bibfield  {journal} {\bibinfo  {journal} {New J. Phys.}\ }\textbf {\bibinfo
  {volume} {13}},\ \bibinfo {pages} {073018} (\bibinfo {year}
  {2011})}\BibitemShut {NoStop}%
\bibitem [{\citenamefont {Moeckel}\ and\ \citenamefont
  {Kehrein}(2008)}]{moeckel08_quench_hubbard_high_d}%
  \BibitemOpen
  \bibfield  {author} {\bibinfo {author} {\bibfnamefont {M.}~\bibnamefont
  {Moeckel}}\ and\ \bibinfo {author} {\bibfnamefont {S.}~\bibnamefont
  {Kehrein}},\ }\href@noop {} {\bibfield  {journal} {\bibinfo  {journal} {Phys.
  Rev. Lett.}\ }\textbf {\bibinfo {volume} {100}},\ \bibinfo {pages} {175702}
  (\bibinfo {year} {2008})}\BibitemShut {NoStop}; %
  \BibitemOpen
  \bibfield  {author} {\bibinfo {author} {\bibfnamefont {M.}~\bibnamefont
  {Moeckel}}\ and\ \bibinfo {author} {\bibfnamefont {S.}~\bibnamefont
  {Kehrein}},\ }\href@noop {} {\bibfield  {journal} {\bibinfo  {journal} {Ann.
  Phys. (N. Y.)}\ }\textbf {\bibinfo {volume} {324}},\ \bibinfo {pages} {2146}
  (\bibinfo {year} {2009})}\BibitemShut {NoStop}%
\bibitem [{\citenamefont {Eckstein}\ \emph {et~al.}(2009)\citenamefont
  {Eckstein}, \citenamefont {Kollar},\ and\ \citenamefont
  {Werner}}]{eckstein09_thermalization_quench_fermi_hubbard}%
  \BibitemOpen
  \bibfield  {author} {\bibinfo {author} {\bibfnamefont {M.}~\bibnamefont
  {Eckstein}}, \bibinfo {author} {\bibfnamefont {M.}~\bibnamefont {Kollar}}, \
  and\ \bibinfo {author} {\bibfnamefont {P.}~\bibnamefont {Werner}},\
  }\href@noop {} {\bibfield  {journal} {\bibinfo  {journal} {Phys. Rev. Lett.}\
  }\textbf {\bibinfo {volume} {103}},\ \bibinfo {pages} {056403} (\bibinfo
  {year} {2009})}\BibitemShut {NoStop}%
\bibitem [{\citenamefont {Kollar}\ \emph {et~al.}(2011)\citenamefont {Kollar},
  \citenamefont {Wolf},\ and\ \citenamefont
  {Eckstein}}]{kollar11_gge_pretherm}%
  \BibitemOpen
  \bibfield  {author} {\bibinfo {author} {\bibfnamefont {M.}~\bibnamefont
  {Kollar}}, \bibinfo {author} {\bibfnamefont {F.~A.}\ \bibnamefont {Wolf}}, \
  and\ \bibinfo {author} {\bibfnamefont {M.}~\bibnamefont {Eckstein}},\
  }\href@noop {} {\bibfield  {journal} {\bibinfo  {journal} {Phys. Rev. B}\
  }\textbf {\bibinfo {volume} {84}},\ \bibinfo {pages} {054304} (\bibinfo
  {year} {2011})}\BibitemShut {NoStop}%
\bibitem [{\citenamefont {Rigol}\ \emph {et~al.}(2007)\citenamefont {Rigol},
  \citenamefont {Dunjko}, \citenamefont {Yurovsky},\ and\ \citenamefont
  {Olshanii}}]{rigol07_generalized_gibbs_hcbosons}%
  \BibitemOpen
  \bibfield  {author} {\bibinfo {author} {\bibfnamefont {M.}~\bibnamefont
  {Rigol}}, \bibinfo {author} {\bibfnamefont {V.}~\bibnamefont {Dunjko}},
  \bibinfo {author} {\bibfnamefont {V.}~\bibnamefont {Yurovsky}}, \ and\
  \bibinfo {author} {\bibfnamefont {M.}~\bibnamefont {Olshanii}},\ }\href@noop
  {} {\bibfield  {journal} {\bibinfo  {journal} {Phys. Rev. Lett.}\ }\textbf
  {\bibinfo {volume} {98}},\ \bibinfo {pages} {050405} (\bibinfo {year}
  {2007})}\BibitemShut {NoStop}%
\bibitem [{\citenamefont {Cazalilla}(2006)}]{cazalilla06_quench_LL}%
  \BibitemOpen
  \bibfield  {author} {\bibinfo {author} {\bibfnamefont {M.~A.}\ \bibnamefont
  {Cazalilla}},\ }\href@noop {} {\bibfield  {journal} {\bibinfo  {journal}
  {Phys. Rev. Lett.}\ }\textbf {\bibinfo {volume} {97}},\ \bibinfo {pages}
  {156403} (\bibinfo {year} {2006})}\BibitemShut {NoStop}; %
  \BibitemOpen
  \bibfield  {author} {\bibinfo {author} {\bibfnamefont {A.}~\bibnamefont
  {Iucci}}\ and\ \bibinfo {author} {\bibfnamefont {M.~A.}\ \bibnamefont
  {Cazalilla}},\ }\href@noop {} {\bibfield  {journal} {\bibinfo  {journal}
  {Phys. Rev. A}\ }\textbf {\bibinfo {volume} {80}},\ \bibinfo {pages} {063619}
  (\bibinfo {year} {2009})}\BibitemShut {NoStop}; %
  N. Nessi and A. Iucci, Phys. Rev. B \textbf{87}, 085137 (2012).%
\bibitem [{\citenamefont {Cazalilla}\ \emph {et~al.}(2012)\citenamefont
  {Cazalilla}, \citenamefont {Iucci},\ and\ \citenamefont
  {Chung}}]{cazalilla12_thermalization_correlations}%
  \BibitemOpen
  \bibfield  {author} {\bibinfo {author} {\bibfnamefont {M.~A.}\ \bibnamefont
  {Cazalilla}}, \bibinfo {author} {\bibfnamefont {A.}~\bibnamefont {Iucci}}, \
  and\ \bibinfo {author} {\bibfnamefont {M.-C.}\ \bibnamefont {Chung}},\
  }\href@noop {} {\bibfield  {journal} {\bibinfo  {journal} {Phys. Rev. E}\
  }\textbf {\bibinfo {volume} {85}},\ \bibinfo {pages} {011133} (\bibinfo
  {year} {2012})}\BibitemShut {NoStop};
   \BibitemOpen
  \bibfield  {author} {\bibinfo {author} {\bibfnamefont {T.}~\bibnamefont
  {Barthel}}\ and\ \bibinfo {author} {\bibfnamefont {U.}~\bibnamefont
  {Schollw\"ock}},\ }\href@noop {} {\bibfield  {journal} {\bibinfo  {journal}
  {Phys. Rev. Lett.}\ }\textbf {\bibinfo {volume} {100}},\ \bibinfo {pages}
  {100601} (\bibinfo {year} {2008})}\BibitemShut {NoStop}%
\bibitem [{\citenamefont {Rigol}\ \emph {et~al.}(2008)\citenamefont {Rigol},
  \citenamefont {Dunjko},\ and\ \citenamefont
  {Olshanii}}]{rigol08_mechanism_thermalization}%
  \BibitemOpen
  \bibfield  {author} {\bibinfo {author} {\bibfnamefont {M.}~\bibnamefont
  {Rigol}}, \bibinfo {author} {\bibfnamefont {V.}~\bibnamefont {Dunjko}}, \
  and\ \bibinfo {author} {\bibfnamefont {M.}~\bibnamefont {Olshanii}},\
  }\href@noop {} {\bibfield  {journal} {\bibinfo  {journal} {Nature (London)}\
  }\textbf {\bibinfo {volume} {452}},\ \bibinfo {pages} {854} (\bibinfo {year}
  {2008})}\BibitemShut {NoStop}%
\bibitem [{Note1()}]{Note1}%
  \BibitemOpen
  \bibinfo {note} {For weak interactions and on dimensional grounds, we expect
  the collision time to be $\sim f_0^{-4} N(0)^{-3}$, where $f_0$ is the
  interaction strength and $N(0)$ is the density of states at the Fermi
  surface, while the prethermalization regime arises for times of the order
  $f_0^{-2} N(0)^{-1}$.}\BibitemShut {Stop}%
\bibitem [{\citenamefont {Lu}\ \emph {et~al.}(2012)\citenamefont {Lu},
  \citenamefont {Burdick},\ and\ \citenamefont {Lev}}]{lu12_erbium_degenerate}%
  \BibitemOpen
  \bibfield  {author} {\bibinfo {author} {\bibfnamefont {M.}~\bibnamefont
  {Lu}}, \bibinfo {author} {\bibfnamefont {N.~Q.}\ \bibnamefont {Burdick}}, \
  and\ \bibinfo {author} {\bibfnamefont {B.~L.}\ \bibnamefont {Lev}},\
  }\href@noop {} {\bibfield  {journal} {\bibinfo  {journal} {Phys. Rev. Lett.}\
  }\textbf {\bibinfo {volume} {108}},\ \bibinfo {pages} {215301} (\bibinfo
  {year} {2012})}\BibitemShut {NoStop}; %
  \BibitemOpen
  \bibfield  {author} {\bibinfo {author} {\bibfnamefont {K.}~\bibnamefont
  {Aikawa}}, \bibinfo {author} {\bibfnamefont {A.}~\bibnamefont {Frisch}},
  \bibinfo {author} {\bibfnamefont {M.}~\bibnamefont {Mark}}, \bibinfo {author}
  {\bibfnamefont {S.}~\bibnamefont {Baier}}, \bibinfo {author} {\bibfnamefont
  {R.}~\bibnamefont {Grimm}}, \ and\ \bibinfo {author} {\bibfnamefont
  {F.}~\bibnamefont {Ferlaino}},\ }\href@noop {} {\bibfield  {journal}
  {\bibinfo  {journal} {arXiv:1310.5676}\ } (\bibinfo {year}
  {2013})}\BibitemShut {NoStop}%
\bibitem [{\citenamefont {Houghton}\ and\ \citenamefont
  {Marston}(1993)}]{houghton93_bosonization_high_d}%
  \BibitemOpen
  \bibfield  {author} {\bibinfo {author} {\bibfnamefont {A.}~\bibnamefont
  {Houghton}}\ and\ \bibinfo {author} {\bibfnamefont {J.~B.}\ \bibnamefont
  {Marston}},\ }\href@noop {} {\bibfield  {journal} {\bibinfo  {journal} {Phys.
  Rev. B}\ }\textbf {\bibinfo {volume} {48}},\ \bibinfo {pages} {7790}
  (\bibinfo {year} {1993})}\BibitemShut {NoStop}; %
  \BibitemOpen
  \bibfield  {author} {\bibinfo {author} {\bibfnamefont {A.}~\bibnamefont
  {Houghton}}, \bibinfo {author} {\bibfnamefont {H.-J.}\ \bibnamefont {Kwon}},
  \ and\ \bibinfo {author} {\bibfnamefont {J.~B.}\ \bibnamefont {Marston}},\
  }\href@noop {} {\bibfield  {journal} {\bibinfo  {journal} {Phys. Rev. B}\
  }\textbf {\bibinfo {volume} {50}},\ \bibinfo {pages} {1351} (\bibinfo {year}
  {1994})}\BibitemShut {NoStop}; %
  \BibitemOpen
  \bibfield  {author} {\bibinfo {author} {\bibfnamefont {A.}~\bibnamefont
  {Houghton}}, \bibinfo {author} {\bibfnamefont {H.-J.}\ \bibnamefont {Kwon}},
  \ and\ \bibinfo {author} {\bibfnamefont {J.~B.}\ \bibnamefont {Marston}},\
  }\href@noop {} {\bibfield  {journal} {\bibinfo  {journal} {Adv. Phys.}\
  }\textbf {\bibinfo {volume} {149}},\ \bibinfo {pages} {141} (\bibinfo {year}
  {2000})}\BibitemShut {NoStop}%
\bibitem [{\citenamefont {{Castro Neto}}\ and\ \citenamefont
  {Fradkin}(1994{\natexlab{a}})}]{castroneto94_bosonization_low_energy_fermi_liquids}%
  \BibitemOpen
  \bibfield  {author} {\bibinfo {author} {\bibfnamefont {A.~H.}\ \bibnamefont
  {{Castro Neto}}}\ and\ \bibinfo {author} {\bibfnamefont {E.~H.}\ \bibnamefont
  {Fradkin}},\ }\href@noop {} {\bibfield  {journal} {\bibinfo  {journal} {Phys.
  Rev. Lett.}\ }\textbf {\bibinfo {volume} {72}},\ \bibinfo {pages} {1393}
  (\bibinfo {year} {1994}{\natexlab{a}})}\BibitemShut {NoStop}; %
  \BibitemOpen
  \bibfield  {author} {\bibinfo {author} {\bibfnamefont {A.~H.}\ \bibnamefont
  {{Castro Neto}}}\ and\ \bibinfo {author} {\bibfnamefont {E.~H.}\ \bibnamefont
  {Fradkin}},\ }\href@noop {} {\bibfield  {journal} {\bibinfo  {journal} {Phys.
  Rev. B}\ }\textbf {\bibinfo {volume} {49}},\ \bibinfo {pages} {10877}
  (\bibinfo {year} {1994}{\natexlab{b}})}\BibitemShut {NoStop}; %
  \BibitemOpen
  \bibfield  {author} {\bibinfo {author} {\bibfnamefont {A.~H.}\ \bibnamefont
  {{Castro Neto}}}\ and\ \bibinfo {author} {\bibfnamefont {E.~H.}\ \bibnamefont
  {Fradkin}},\ }\href@noop {} {\bibfield  {journal} {\bibinfo  {journal} {Phys.
  Rev. B}\ }\textbf {\bibinfo {volume} {51}},\ \bibinfo {pages} {4084}
  (\bibinfo {year} {1995})}\BibitemShut {NoStop}%
\bibitem [{\citenamefont {Karrasch}\ \emph {et~al.}(2012)\citenamefont
  {Karrasch}, \citenamefont {Rentrop}, \citenamefont {Schuricht},\ and\
  \citenamefont {Meden}}]{karrasch12_ll_universality_quench}%
  \BibitemOpen
  \bibfield  {author} {\bibinfo {author} {\bibfnamefont {C.}~\bibnamefont
  {Karrasch}}, \bibinfo {author} {\bibfnamefont {J.}~\bibnamefont {Rentrop}},
  \bibinfo {author} {\bibfnamefont {D.}~\bibnamefont {Schuricht}}, \ and\
  \bibinfo {author} {\bibfnamefont {V.}~\bibnamefont {Meden}},\ }\href@noop {}
  {\bibfield  {journal} {\bibinfo  {journal} {Phys. Rev. Lett.}\ }\textbf
  {\bibinfo {volume} {109}},\ \bibinfo {pages} {126406} (\bibinfo {year}
  {2012})}\BibitemShut {NoStop}; %
  \BibitemOpen
  \bibfield  {author} {\bibinfo {author} {\bibfnamefont {J.}~\bibnamefont
  {Rentrop}}, \bibinfo {author} {\bibfnamefont {D.}~\bibnamefont {Schuricht}},
  \ and\ \bibinfo {author} {\bibfnamefont {V.}~\bibnamefont {Meden}},\
  }\href@noop {} {\bibfield  {journal} {\bibinfo  {journal} {New J. Phys.}\
  }\textbf {\bibinfo {volume} {14}},\ \bibinfo {pages} {075001} (\bibinfo
  {year} {2012})}\BibitemShut {NoStop}%
\bibitem{mitra11_quench_mode_coupling} A. Mitra and T. Giamarchi, Phys. Rev. Lett. \textbf{107}, 150602 (2011).
\bibitem [{sup()}]{supplementary}%
  \BibitemOpen
  \href@noop {} {\ }\bibinfo {note} {See supplementary material}\BibitemShut
  {NoStop}%
\bibitem [{\citenamefont {Kopietz}(1997)}]{kopietz99_book_bosonization_high_D}%
  \BibitemOpen
  \bibfield  {author} {\bibinfo {author} {\bibfnamefont {P.}~\bibnamefont
  {Kopietz}},\ }\href@noop {} {\emph {\bibinfo {title} {Bosonization of
  Interacting Fermions in Arbitary Dimensions}}}\ (\bibinfo  {publisher}
  {Springer},\ \bibinfo {year} {1997})\BibitemShut {NoStop}%
\bibitem [{\citenamefont {Calabrese}\ and\ \citenamefont
  {Cardy}(2006)}]{calabrese06_quench_CFT}%
  \BibitemOpen
  \bibfield  {author} {\bibinfo {author} {\bibfnamefont {P.}~\bibnamefont
  {Calabrese}}\ and\ \bibinfo {author} {\bibfnamefont {J.}~\bibnamefont
  {Cardy}},\ }\href@noop {} {\bibfield  {journal} {\bibinfo  {journal} {Phys.
  Rev. Lett.}\ }\textbf {\bibinfo {volume} {96}},\ \bibinfo {pages} {136801}
  (\bibinfo {year} {2006})}\BibitemShut {NoStop}%
\end{thebibliography}


\clearpage

\begin{widetext}

\section{Supplementary material to ``Quantum Quench and Prethermalization Dynamics in A Two-Dimensional Fermi Gas
with Long-range Interactions''}

\setcounter{equation}{0}

\subsection{Truncation of the Hamiltonian and bosonization}

The Fermi surface (FS) bosonization is a technique that allows to deal with the low-energy degrees of freedom of a Fermi system. In equilibrium, the low-energy degrees of freedom dominate the
low-temperature, long-wave length, low frequency properties
and response to external probes of the system.
The effective description of the low-energy degrees
of freedom, known as Landau Fermi liquid theory, is
obtained after coarse graining the high-energy fluctuations
of the system by means of a renormalization group procedure.

 In our case such a coarse-graining does not apply as we are interested in the short to intermediate time dynamics of an interacting many body system which is driven out of equilibrium.  Thus,  in order to work only with the low energy degrees of freedom, we must appeal to the observation made in the main text that for long-range interactions (i.e. those for which the characteristic momentum transfer is smaller than the Fermi momentum, $k_F$) and short to intermediate times after the interaction is quenched, the production of particle-hole pairs with small momentum dominates the dynamics.
Furthermore, we also assume that inelastic collisions
between the excitations set in at much longer times, and
therefore the initial response of the system can be
descried by forward and exchange interactions.

 In order to apply the FS bosonization machinery to the interaction quench, we start from the  Hamiltonian of a Fermi gas with
two-body interactions:
\begin{equation}\label{eq:fermion_ham}
H=H_0+H_{int}=\sum_{\mathbf{k}}\epsilon(\mathbf{k})c^{\dagger}_{\mathbf{k}}c_{\mathbf{k}}+\frac{1}{2V}\sum_{\mathbf{k}_1,\mathbf{k}_2,\mathbf{q}}f(q)J_{\mathbf{k}_1}(\mathbf{q})J_{\mathbf{k}_2}(-\mathbf{q}),
\end{equation}
where $\{c_{\mathbf{k}},c^{\dagger}_{\mathbf{k}}\}$ are the fermionic field operators and $J_{\mathbf{k}}(\mathbf{q})=c^{\dagger}_{\mathbf{k}+\mathbf{q}}c_{\mathbf{k}}$ create particle-hole excitations. In this expression the momentum sums over $\mathbf{k}_1$, $\mathbf{k}_2$ and $\mathbf{q}$ run over the entire range of momentum values. 
In order to implement the FS bosonization we introduce the patch construction: We divide the FS into $N$ patches of size $\Lambda$ tangent to the FS, and label the patches by the index $\mathbf{S}$, such that $\mathbf{k}_{\mathbf{S}}$ denotes the vector right at the FS in the middle of the patch. Let us assume that the typical momentum transfer of the interaction is $q_c$, we choose the radial cutoff $\lambda$ such that $q_c<\lambda$. We introduce the functions
\begin{equation}
\Theta(\mathbf{S};\mathbf{p})= \begin{cases} 1 &\mbox{if } \mathbf{p} \in \mbox{patch $\mathbf{S}$} \\
0 & \mbox{if } \mbox{otherwise}. \end{cases}
\end{equation}
We also introduce auxiliary functions $\tilde{\Theta}(\mathbf{S};\mathbf{p})$ that are zero for $\mathbf{p}$ in patch $\mathbf{S}$ and non-zero in a complimentary region of that patch, such that
\begin{equation}\label{eq:completeness}
\sum_{\mathbf{S}}\Theta(\mathbf{S};\mathbf{p})+\tilde{\Theta}(\mathbf{S};\mathbf{p})=1,
\end{equation}
meaning that the superposition of the patches and its complimentary regions cover the whole momentum space. In order to isolate the degrees of freedom that are close to the FS, we introduce the completeness relation Eq. (\ref{eq:completeness}) repeatedly in the interacting part of the fermionic Hamiltonian (\ref{eq:fermion_ham}) and group terms in the following way
\begin{equation}\label{eq:truncated_int}
H_{int}=\frac{1}{2V}\sum_{\mathbf{S},\mathbf{T},\mathbf{S}',\mathbf{T}',\mathbf{q}}f(q)\Theta(\mathbf{S};\mathbf{k}_1)\Theta(\mathbf{S}';\mathbf{k}_1+\mathbf{q})\Theta(\mathbf{T};\mathbf{k}_2)\Theta(\mathbf{T}';\mathbf{k}_2-\mathbf{q})J_{\mathbf{k}_1}(\mathbf{q})J_{\mathbf{k}_2}(-\mathbf{q})+\tilde{H}_{int},
\end{equation}
where $\tilde{H}_{int}$ contains at least one $\tilde{\Theta}$, i.e., is the Hamiltonian that involves the high-energy degrees of freedom. Our main approximation is to neglect $\tilde{H}_{int}$ in the description of the short time dynamics. Basically, these amounts to neglect profoundly inelastic processes that, as argued before, will become relevant only on a second stage of the relaxation. A careful consideration of Eq. (\ref{eq:truncated_int}) reveals that there exist only two types of non-zero terms in the sum, $\mathbf{S}=\mathbf{S}'$ and $\mathbf{T}=\mathbf{T}'$ (forward scattering) or $\mathbf{S}=\mathbf{T}'$ and $\mathbf{T}=\mathbf{S}'$ (exchange scattering). Introducing the coarse-grained densities
\begin{equation}
J_{\mathbf{S}}(\mathbf{q})=\sum_{\mathbf{k}}\Theta(\mathbf{S};\mathbf{k}+\mathbf{q})\Theta(\mathbf{S};\mathbf{k})J_{\mathbf{k}}(\mathbf{q}),
\end{equation}
the forward scattering terms can be immediately written as
\begin{equation}\label{eq:boson_int}
H_{int}=\sum_{\mathbf{S},\mathbf{T},\mathbf{q}}f(q)J_{\mathbf{S}}(\mathbf{q})J_{\mathbf{T}}(-\mathbf{q}),
\end{equation}
where only vectors $\mathbf{q}$ that are small enough to fit into one patch are allowed. In general, the exchange terms cannot be written in terms of the coarse-grained densities. However they can be absorbed into the forward scattering part if we consider that the size $\Lambda$ of the patch is much larger than the interaction cutoff $q_c$. In such case, if the interaction is not too strong, the exchange terms that transfer momentum from one patch to another will be accompanied by a tiny matrix element $f(q)$ and we can neglect them as well. The only exchange terms left are those with $\mathbf{S}=\mathbf{T}=\mathbf{S}'=\mathbf{T}'$ which are indistinguishable of the diagonal terms of the forward scattering terms.

The main step of the bosonization procedure is to note that the in-patch densities obey the anomalous commutation relations
\begin{equation}\label{eq:conmut}
[J_{\mathbf{S}}(\mathbf{q}),J_{\mathbf{T}}(\mathbf{p})]=\delta_{\mathbf{S},\mathbf{T}}\delta_{\mathbf{q}+\mathbf{p}}\,\Omega\,\mathbf{\hat{n}_S}\cdot\mathbf{q}+\mathrm{Error}.
\end{equation}
In general, the error term is small if $\Lambda\gg \lambda$ which determines a squat aspect ratio for the FS patches. If we neglect the error term, it is possible to write the currents in terms of canonical bosonic operators $\{a_{\mathbf{S}}(\mathbf{q}),a^{\dagger}_{\mathbf{S}}(\mathbf{q})\}$:
\begin{equation}
J_{\mathbf{S}}(\mathbf{q})=\sqrt{\Omega\vert \mathbf{\hat{n}_S}\cdot\mathbf{q}\vert}[a^{\dagger}_{\mathbf{S}}(\mathbf{q})\theta(\mathbf{\hat{n}_S}\cdot\mathbf{q})+a_{\mathbf{S}}(-\mathbf{q})\theta(-\mathbf{\hat{n}_S}\cdot\mathbf{q})],
\end{equation}
where $\theta(x)$ is the Heaviside function. Then, the low-energy part of the interacting Hamiltonian Eq. (\ref{eq:boson_int}) is already bosonized. To bosonize the low-energy part of the kinetic Hamiltonian we have to made further approximations. Starting from the kinetic energy
\begin{align}\label{eq:kin_teunc}
H_0=\sum_{\mathbf{S},\mathbf{q}}\epsilon(\mathbf{k}_{\mathbf{S}}+\mathbf{q})c^{\dagger}_{\mathbf{k}_{\mathbf{S}}+\mathbf{q}}c_{\mathbf{k}_{\mathbf{S}}+\mathbf{q}}+\tilde{H}_0,
\end{align}
we again neglect $\tilde{H}_0$. Moreover, we assume that the patch size $\Lambda$ is small enough compared to the scale in which the FS changes its shape. For the circular FS is enough to ask $\Lambda\ll k_F$. In this approximation we can neglect the variations of the vector normal to the FS inside each patch, making possible to linearize the dispersion relation inside each patch~\cite{kopietz99_book_bosonization_high_D}. Additionally, if we focus only on the low energy degrees of freedom, in virtue of the commutation relations (\ref{eq:conmut}), we can write the low-energy kinetic part of the hamiltonian as~\cite{castroneto94_bosonization_low_energy_fermi_liquids}
\begin{equation}
\frac{v_F}{2\Omega}\sum_{\mathbf{S},\mathbf{q}}J_{\mathbf{S}}(\mathbf{q}) J_{\mathbf{S}}(-\mathbf{q}),
\end{equation}
where $v_F=\vert\nabla_{\mathbf{k}}\epsilon(\mathbf{k})\vert_{\vert\mathbf{k}\vert=k_F}$ is the Fermi velocity. The bosonized Hamiltonian can be finally written as
\begin{equation}\label{eq:ham}
H=\frac{1}{2}\sum_{\mathbf{S},\mathbf{T},\mathbf{q}}J_{\mathbf{S}}(\mathbf{q})\left(\frac{v_F}{\Omega} \delta_{\mathbf{S},\mathbf{T}} +\frac{f(q)}{V}\right)J_{\mathbf{T}}(-\mathbf{q}).
\end{equation}
One important thing to notice is that the interaction potential $f(q)$ that appears in the bosonized interaction Hamiltonian above is the \emph{bare} interaction between the fermions. In contrast, in the equilibrium case this interaction parameter is the renormalized interaction between the quasiparticles that remain after the high-energy degrees of freedom had been integrated out.

The collection of conditions over the cutoffs is consistent and reads:
\begin{equation}
k_F\gg\Lambda\gg\lambda>q_c.
\end{equation}

\subsection{Solution of the dynamics}

In terms of the bosonic operators the truncated Hamiltonian can be written as
\begin{equation}\label{eq:bos_ham}
H=\sum_{\mathbf{q},\mathbf{S}\mathbf{T}}\Big[a^{\dagger}_{\mathbf{S}}(\mathbf{q})V_{\mathbf{S}\mathbf{T}}(\mathbf{q})a_{\mathbf{T}}(\mathbf{q})+\left( a^{\dagger}_{\mathbf{S}}(\mathbf{q})W_{\mathbf{S}\mathbf{T}}(\mathbf{q})a^{\dagger}_{\mathbf{T}}(-\mathbf{q})+\mathrm{H.c.}\right)\Big],
\end{equation}
where $V_{\mathbf{S}\mathbf{T}}$ and $W_{\mathbf{S}\mathbf{T}}$
describe the coupling within and between different patches resulting
from the kinetic and interaction terms in~\eqref{eq:ham}. They are given by
\begin{align}
V_{\mathbf{S}\mathbf{T}} =&\,\delta_{\mathbf{S}\mathbf{T}} 2\theta(\hat{\mathbf{n}}_{\mathbf{S}}\cdot\mathbf{q}) \hat{\mathbf{n}}_{\mathbf{S}}\cdot\mathbf{q}\nonumber+ \frac{2 f(q)\Lambda}{(2\pi)^2} \theta(\hat{\mathbf{n}}_{\mathbf{S}}\cdot\mathbf{q}) \theta(\hat{\mathbf{n}}_{\mathbf{T}}\cdot\mathbf{q})  \sqrt{(\hat{\mathbf{n}}_{\mathbf{S}}\cdot\mathbf{q})(\hat{\mathbf{n}}_{\mathbf{T}}\cdot\mathbf{q})}, \\
W_{\mathbf{S}\mathbf{T}} =&\,\frac{f(q)\Lambda}{(2\pi)^2} \theta(\hat{\mathbf{n}}_{\mathbf{S}}\cdot\mathbf{q})\theta(-\hat{\mathbf{n}}_{\mathbf{T}}\cdot\mathbf{q})\sqrt{\left(\hat{\mathbf{n}}_{\mathbf{S}}\cdot\mathbf{q}\right)\left(-\hat{\mathbf{n}}_{\mathbf{T}}\cdot\mathbf{q}\right)}.
\end{align}
The initial state is the ground state of the Hamiltonian that is obtained from $H$ by setting $f(q)=0$, i.e., the vacuum of the $a_{\mathbf{S}}(\mathbf{q})$ operators. The general form of the solution to the Heisenberg equations of motion for $a_{\mathbf{S}}$ takes the form
\begin{equation}
e^{iHt}a_{\mathbf{S}}(\mathbf{q})e^{-iHt}=\sum_{\mathbf{T}}A_{\mathbf{ST}}(t)a_{\mathbf{T}}(\mathbf{q})+B_{\mathbf{ST}}(t)a^{\dagger}_{\mathbf{T}}(-\mathbf{q})
\end{equation}
where the matrices $A_{\mathbf{ST}}(t)$ and $B_{\mathbf{ST}}(t)$ can be obtained by using a canonical transformation for the diagonalization of $H$~\cite{cazalilla06_quench_LL}. We could try to proceed in this way by using the explicit form of the transformation~\cite{castroneto94_bosonization_low_energy_fermi_liquids}. However this approach turns out to be a bit cumbersome for the out of equilibrium evolution. Instead, we shall follow an alternative route by noticing that matrices $A_{\mathbf{SR}}(t)$ and $B_{\mathbf{SR}}(t)$ can be written as commutators between time-evolved and static operators:
\begin{align}\label{eq:heisenberg}
\nonumber A_{\mathbf{ST}}(t) &= [a_{\mathbf{S}}(\mathbf{q},t),a^{\dagger}_{\mathbf{T}}(\mathbf{q})],\\
B_{\mathbf{ST}}(t) &= [a_{\mathbf{S}}(\mathbf{q},t),a_{\mathbf{T}}(-\mathbf{q})].
\end{align}
Since the Hamiltonian is quadratic, these commutators are just c-numbers, and therefore we are free to take their expectation value in any normalized state. As the operators evolve according to $H$ we choose its ground state $\vert\Psi\rangle$ to perform the average. Therefore these matrices can be written for $t>0$ in terms of the following \emph{equilibrium} retarded correlation functions
\begin{align}\label{eq:heisenberg}
\nonumber G^{\mathrm{ret,eq}}_{\mathbf{S}\mathbf{T}}(\mathbf{q},t) &= -i\theta(t)\langle\Psi\vert[a_{\mathbf{S}}(\mathbf{q},t),a^{\dagger}_{\mathbf{T}}(\mathbf{q})]\vert\Psi\rangle,\\
\tilde{G}^{\mathrm{ret,eq}}_{\mathbf{S}\mathbf{T}}(\mathbf{q},t) &= -i\theta(t)\langle\Psi\vert[a_{\mathbf{S}}(\mathbf{q},t),a_{\mathbf{T}}(-\mathbf{q})]\vert\Psi\rangle.
\end{align}
The (Fourier transformed) retarded functions can be calculated by analytic continuation of the Matsubara functions, which in turns are obtained within the path integral formulation of the equilibrium bosonic problem of Eq. (\ref{eq:bos_ham}) by means of a Hubbard-Stratonovich transformation~\cite{houghton93_bosonization_high_d}. In this way we avoid to use the explicit form of the generalized Bogoliubov transformation that diagonalizes the Hamiltonian~\cite{castroneto94_bosonization_low_energy_fermi_liquids}.

\subsection{Calculation of the Fermionic Green's function}

The non-equilibrium one particle density matrix is defined through
\begin{equation}
\mathcal{G}^{\mathrm{neq}}(\mathbf{x},t)=\langle\Psi_0\vert e^{iHt}\psi^{\dagger}(\mathbf{x})\psi(\mathbf{0})e^{iHt}\vert\Psi_0\rangle,
\end{equation}
which can be written as a sum over patches,
$\mathcal{G}^{\mathrm{neq}}(\mathbf{x},t)=\sum_{\mathbf{S}}\mathcal{G}^{\mathrm{neq}}_{\mathbf{S}}(\mathbf{x},t)$, with
\begin{equation}
\mathcal{G}^{\mathrm{neq}}_{\mathbf{S}}(\mathbf{x},t)
=\langle\Psi_0\vert e^{iHt}\psi^{\dagger}_{\mathbf{S}}(\mathbf{x})\psi_{\mathbf{S}}(\mathbf{0})e^{iHt}\vert\Psi_0\rangle,
\end{equation}
and where $\psi_{\mathbf{S}}(\mathbf{x})$ is the fermion field operator that annihilates a fermion at position $\mathbf{x}$ with momenta within the patch $\mathbf{S}$. The calculation can be carried out by means of the bosonic representation of the fermionic field~\cite{houghton93_bosonization_high_d,castroneto94_bosonization_low_energy_fermi_liquids}
\begin{equation}
\psi_{\mathbf{S}}(\mathbf{x})=e^{i\mathbf{k_S}\cdot\mathbf{x}}\sqrt{\frac{\Omega}{Va}}\exp\left[i\frac{1}{\sqrt{\Omega}}\phi_{\mathbf{S}}(\mathbf{x})\right], \end{equation}
where $a=1/\lambda$ is the ultraviolet cutoff and the bosonic field  $\phi_{\mathbf{S}}(\mathbf{x})=i\sum_{\mathbf{q},\mathbf{\hat{n}_S}\cdot\mathbf{q}>0}\frac{a^{\dagger}_{\mathbf{S}}(\mathbf{q})e^{-i\mathbf{q}\cdot \mathbf{x}}-a_{\mathbf{S}}(\mathbf{q})e^{i\mathbf{q}\cdot \mathbf{x}}}{\sqrt{\mathbf{\hat{n}_S}\cdot\mathbf{q}}}$. In order to respect the initial condition, the correlation function must admit the factorization $\mathcal{G}^{\mathrm{neq}}_{\mathbf{S}}(\mathbf{x},t)=\mathcal{G}^{0}_{\mathbf{S}}(\mathbf{x})Z^{\mathrm{neq}}_{\mathbf{S}}(\mathbf{x},t)$, where
\begin{equation}
\mathcal{G}^{0}_{\mathbf{S}}(\mathbf{x})=\frac{i\Lambda}{(2\pi)^{2}}\frac{e^{ik_{F}\mathbf{\hat{n}}_{\mathbf{S}}\cdot \mathbf{x}}}{\mathbf{\hat{n}}_{\mathbf{S}}\cdot \mathbf{x}}W_{\Lambda}(\vert\mathbf{\hat{n}}_{\mathbf{S}}\times \mathbf{x}\vert)
\end{equation}
is the equilibrium correlation function in the initial (free) state, and $Z^{\mathrm{neq}}_{\mathbf{S}}(\mathbf{x},0)=1$. $W_{\Lambda}(\vert\mathbf{\hat{n}}_{\mathbf{S}}\times \mathbf{x}\vert)$ is a windowing function that vanishes for $\vert\mathbf{\hat{n}}_{\mathbf{S}}\times \mathbf{x}\vert\Lambda\gtrsim 1$ and attains the value of unity for $\vert\mathbf{\hat{n}}_{\mathbf{S}}\times \mathbf{x}\vert=0$. Using the solution (\ref{eq:heisenberg}) we find:
\begin{equation}\label{eq:z_ret_supp}
Z^{\mathrm{neq}}_{\mathbf{S}}(\mathbf{x},t)=\exp\Bigg[\frac{1}{\Omega}\sum_{\mathbf{q},\mathbf{\hat{n}_S}\cdot\mathbf{q}>0,\mathbf{T}}\frac{\vert\tilde{G}^{\mathrm{ret,eq}}_{\mathbf{S}\mathbf{T}}(\mathbf{q},t)\vert^2}{\mathbf{\hat{n}_S}\cdot\mathbf{q}}  2(\cos(\mathbf{q}\cdot\mathbf{x})-1)\Bigg],
\end{equation}
where
\begin{equation}\label{eq:g_ret_supp}
\vert\tilde{G}^{\mathrm{ret,eq}}_{\mathbf{S}\mathbf{T}}(\mathbf{q},t)\vert^2=\frac{2f^{2}(q)\Lambda^{2}}{(2\pi)^{4}}\mathcal{PV}\frac{\vert\mathbf{\hat{n}_S}\cdot\mathbf{q}\vert\vert\mathbf{\hat{n}_T}\cdot\mathbf{q}\vert}{\left(\mathbf{\hat{n}_S}\cdot\mathbf{q}-\mathbf{\hat{n}_T}\cdot\mathbf{q}\right)^{2}}(1-\cos[(\mathbf{\hat{n}_S}\cdot\mathbf{q}-\mathbf{\hat{n}_T}\cdot\mathbf{q})t])+\mathcal{O}(f_{0}^{3}),
\end{equation}
where $\mathcal{PV}$ indicates that in the limit of large number of patches, where patch sums can be approximated by angular integrals, we must take the principal value, a mathematical feature which is also present in the equilibrium calculations~\cite{castroneto94_bosonization_low_energy_fermi_liquids}. In the last line we show only the lowest order in the interaction strength. For large times, the oscillatory factor in Eq.~(\ref{eq:g_ret_supp}) drops out.

The equilibrium's Green's function can also be calculated using the Matsubara formalism. We obtain:
\begin{equation}\label{eq:z_eq}
Z^{\mathrm{eq}}_{\mathbf{S}}(\mathbf{x})=\exp\Bigg[\frac{1}{\Omega}\sum_{\mathbf{q},\mathbf{\hat{n}_S}\cdot\mathbf{q}>0,\mathbf{T}}  \frac{2
f(q)\Lambda^{2}}{(2\pi)^{4}}\mathcal{PV}\frac{\vert\mathbf{\hat{n}_S}\cdot\mathbf{q}\vert\vert\mathbf{\hat{n}_T}\cdot\mathbf{q}\vert}{\left(\mathbf{\hat{n}_S}\cdot\mathbf{q}-\mathbf{\hat{n}_T}\cdot\mathbf{q}\right)^{2}} 2(\cos(\mathbf{q}\cdot\mathbf{x})-1)+\mathcal{O}(f_{0}^{3})\Bigg].
\end{equation}
Due to the presence of the function $W_{\Lambda}$ and of the exponential factor $e^{ik_{F}\mathbf{\hat{n}}_{\mathbf{S}}\cdot\mathbf{x}}$ in $\mathcal{G}^{0}_{\mathbf{S}}(\mathbf{x})$, the final patch sum needed to obtain the full fermion correlator (at equilibrium and after the quench) is dominated by contributions coming from patches where $\phi_{\mathbf{S}}=\arccos(\frac{\mathbf{\hat{n}}_{\mathbf{S}}\cdot \mathbf{x}}{\vert\mathbf{x}\vert})$ is close to zero. Moreover, since $Z^{\mathrm{neq}}_{\mathbf{S}}(\mathbf{x},t)$ varies slowly around $\phi_{\mathbf{S}}=0$, the global correlation function can be accurately approximated as
\begin{equation}\label{eq:corr_global}
\mathcal{G}^{\mathrm{neq}}(x,t)=\mathcal{G}^{0}(x)Z^{\mathrm{neq}}(x,t),
\end{equation}
where $\mathcal{G}^{0}(x)$ is the bare correlator, $x=\vert\mathbf{x}\vert$ and $Z^{\mathrm{neq}}(x,t)=Z^{\mathrm{neq}}_{\mathbf{S}}(\mathbf{x},t)\mid_{\phi_{\mathbf{S}}=0}$, and an analogous expression for the equilibrium's Green's function. Comparing Eq. (\ref{eq:z_eq}) with Eq. (\ref{eq:z_ret_supp}) leads to Eq. (\ref{eq:z_neq}) in the main text.

\subsubsection{Light-cone effect}

Starting from Eqs.~\eqref{eq:z_ret_supp},\eqref{eq:g_ret_supp}, we can show that there exists a light-cone effect in the correlations. First we notice that
\begin{align}
\nonumber(\cos(\mathbf{q}\cdot\mathbf{x})-1)(1-\cos[(\mathbf{\hat{n}_S}\cdot\mathbf{q}-\mathbf{\hat{n}_T}\cdot\mathbf{q})t])=&-\frac{1}{2}\{\cos[\mathbf{q}\cdot(\mathbf{x}+(\mathbf{\hat{n}_S}-\mathbf{\hat{n}_T})t)]+\cos[\mathbf{q}\cdot(\mathbf{x}-(\mathbf{\hat{n}_S}-\mathbf{\hat{n}_T})t)]\}\\
&-1+\cos(\mathbf{q}\cdot\mathbf{x})+\cos[(\mathbf{\hat{n}_S}\cdot\mathbf{q}-\mathbf{\hat{n}_T}\cdot\mathbf{q})t].
\end{align}
Then, if $\vert\mathbf{x}\vert\gg \vert\mathbf{\hat{n}_S}-\mathbf{\hat{n}_T}\vert v_F t$ for all $\mathbf{S}$ and $\mathbf{T}$, i.e., if $\vert\mathbf{x}\vert\gg 2v_F t$, then, at first order in $\frac{2v_F t}{\vert\mathbf{x}\vert}$, we can neglect the spatial dependence of $Z^{\mathrm{neq}}_{\mathbf{S}}(\mathbf{x},t)$. It is important to note, then, that after the momentum integrals in Eq. (\ref{eq:z_ret_supp}) are performed, the patch index of $Z^{\mathrm{neq}}_{\mathbf{S}}(\mathbf{x},t)$ is only preserved through its spatial dependence. Outside the light cone, $\vert\mathbf{x}\vert\gg 2v_Ft$, the interaction correction is therefore approximately the same for all patches: $Z^{\mathrm{neq}}_{\mathbf{S}}(\mathbf{x},t)\approx Z^{\mathrm{neq}}(t)$.
The full correlation function thus reads $\mathcal{G}^{\mathrm{neq}}(x,t)\approx\mathcal{G}^{\mathrm{0}}(x)Z^{\mathrm{neq}}(t)$, i.e., the correlations retain the same spatial dependence as in the initial state up to a time-dependent prefactor. This factor defines the time-dependent quasiparticle residue. On the opposite limit, $\vert\mathbf{x}\vert \ll 2v_F t$, we can neglect the temporal dependence and the steady state correlations dominate: $Z^{\mathrm{neq}}(x,t)\approx \lim_{t\rightarrow\infty}Z^{\mathrm{neq}}(x,t)$.
\end{widetext}

\end{document}